\documentclass[runningheads]{llncs}
\usepackage[T1]{fontenc}
\usepackage{graphicx}
\usepackage{amsmath,amsfonts}
\usepackage{algorithmic}
\usepackage{algorithm}
\usepackage{array}
\usepackage{textcomp}
\usepackage{stfloats}
\usepackage{url}
\usepackage{verbatim}
\usepackage{graphicx}
\usepackage{cite}
\usepackage{multirow} 
\usepackage{float}
\graphicspath{ {image/} }

\usepackage{url}
\usepackage{makecell}
\usepackage{multirow}
\usepackage{rotating}
\usepackage{threeparttable} 
\usepackage{amsmath}  
\usepackage{amsfonts,amssymb}

\usepackage{booktabs}
\usepackage{arydshln}
\usepackage[misc]{ifsym}
\usepackage{xcolor}

\begin{document}
\title{Dual-view Aware Smart Contract Vulnerability Detection for Ethereum}
%
%
\author{
Jiacheng Yao\inst{1,2} \and
Maolin Wang\inst{1,2} \and
Wanqi Chen\inst{1,2} \and
Chengxiang Jin\inst{1,2} \and \\
Jiajun Zhou\inst{1,2} \textsuperscript{(\Letter)} \and
Shanqing Yu\inst{1,2} \and
Qi Xuan\inst{1,2}
}

\authorrunning{Yao et al.}
%
\institute{
Institute of Cyberspace Security, Zhejiang University of Technology, \\Hangzhou 310023, China  
\and
Binjiang Institute of Artificial Intelligence, ZJUT, \\Hangzhou 310056, China \\
\email{jjzhou@zjut.edu.cn}}
%
\maketitle              
\begin{abstract}
The wide application of Ethereum technology has brought technological innovation to traditional industries. As one of Ethereum's core applications, smart contracts utilize diverse contract codes to meet various functional needs and have gained widespread use. However, the non-tamperability of smart contracts, coupled with vulnerabilities caused by natural flaws or human errors, has brought unprecedented challenges to blockchain security. Therefore, in order to ensure the healthy development of blockchain technology and the stability of the blockchain community, it is particularly important to study the vulnerability detection techniques for smart contracts. In this paper, we propose a \textbf{D}ual-view Aware Smart Contract \textbf{V}ulnerability \textbf{Det}ection Framework named DVDet. The framework initially converts the source code and bytecode of smart contracts into weighted graphs and control flow sequences, capturing potential risk features from these two perspectives and integrating them for analysis, ultimately achieving effective contract vulnerability detection. Comprehensive experiments on the Ethereum dataset show that our method outperforms others in detecting vulnerabilities.

\end{abstract}

\section{Introduction}
Ethereum~\cite{wood2014ethereum} is the first blockchain platform to achieve Turing completeness, enabling a variety of functional applications through smart contracts. Smart contracts~\cite{zheng2020overview}, one of Ethereum's core technologies, define the rules and conditions of contracts in the form of programming code on the blockchain, enabling automatic contract execution. These contracts can automatically handle various tasks such as asset transfers, voting execution, and digital asset management without relying on third-party trust mechanisms. However, smart contracts also pose potential risks due to vulnerabilities that may lead to stolen funds, abnormal contract behavior, or data leaks, causing financial losses and trust crises for users~\cite{mense2018security}. For example, incidents like TheDao event and the Poly Network attack were triggered by smart contract vulnerabilities, leading to significant financial losses. Since 2016, the number of blockchain security incidents has been increasing annually, with a significant proportion caused by smart contract vulnerabilities. These incidents not only jeopardize the property safety of users but also intensify public skepticism about the security of blockchain platforms, prompting ongoing research into smart contract security.

Existing methods for detecting vulnerabilities in smart contracts include static analysis, dynamic analysis, symbolic execution, fuzz testing, and deep learning-based approaches.
However, each of them has its limitations. 
For instance, static analysis cannot capture the dynamic behaviors of contracts, while dynamic analysis might not cover all execution paths. Symbolic execution and fuzz testing are affected by the path explosion problem, and in terms of accuracy, existing methods often result in false positives or false negatives. Moreover, certain detection techniques may only be applicable to specific types of vulnerabilities and lack effective means to detect new or complex vulnerabilities.

To address these issues, we propose a \textbf{D}ual-view Aware Smart Contract \textbf{V}ulnerability \textbf{Det}ection Framework (DVDet), which focuses on both source code and bytecode to detect vulnerabilities in smart contracts.
For source code view, it constructs an augmented contract code graph and feed it to a improved graph neural network model to capture the inherent logical semantics within the code. 
For bytecode view, it constructs a control flow sequence and enhances the sequence model to extract thorough sequence features.
Finally, it integrates the features from both views to achieve effective detection for smart contract vulnerabilities. 
The main contributions of this paper can be summarized as follows:
\begin{itemize}
    \item[$\bullet$] We propose a unique Dual-view Aware Smart Contract Vulnerability Detection Framework, which combines features from both the source code and bytecode of smart contracts for vulnerability detection.
    \item[$\bullet$] We propose a data augmentation method for abstract syntax trees by quantifying node importance to further assign edge weights, ultimately converting the augmented abstract syntax tree into a weighted smart contract graph.
    \item[$\bullet$] We propose the HyperAGRU model, which integrates attention mechanisms into GRU units. This not only enhances the ability to capture local features of control flow sequences but also effectively highlights the crucial information inherent in the control flow.
    \item[$\bullet$] Extensive experiments on real-world datasets demonstrate that our DVDet outperforms existing models for smart contract vulnerability detection.
\end{itemize}

\section{Related Work}
\label{related work}
\subsection{Traditional Vulnerability Detection Methods}
Traditional vulnerability detection usually relies on analyzing the underlying logic of the contract.

First, static analysis detects vulnerabilities by analyzing source code syntax, semantics, and data flow.
This method can identify issues like reentrancy attacks, overflows, and uninitialized variables, but it is weak to detect complex vulnerabilities.
Feist et al.~\cite{feist2019slither} detect vulnerabilities by converting Ethereum smart contract code into an intermediate representation.
Schneidewind et al.~\cite{schneidewind2020ethor} perform reachability analysis based on EVM bytecode to detect potential vulnerabilities.
Second, dynamic analysis detects vulnerabilities by simulating different transactions and operations during actual contract execution. 
It can capture dynamic information during program runtime, such as memory usage, but requires more resources and time. 
Azzopardi et al.~\cite{azzopardi2018monitoring} use dynamic event automata to monitor the control flow and data flow events of contracts. Wustholz et al.~\cite{wustholz2020harvey} propose a lightweight gray-box fuzz testing tool mainly used to detect common vulnerability types. 
Third, symbolic execution explores contract paths through symbolic variables to discover potential vulnerabilities. It excels at uncovering subtle errors and complex issues but is susceptible to path explosion and has limited capabilities in handling complex data structures. 
Veloso et al.~\cite{veloso2021conkas} combine the advantages of static analysis and symbolic execution to simulate contract execution paths under different input conditions.
Finally, fuzz testing observes contract behavior by randomly generating inputs to find anomalies. It can discover uncommon boundary cases and anomalies but may generate a large number of invalid inputs, leading to false positives. 
Nguyen et al.~\cite{nguyen2020sfuzz} improve testing efficiency using an adaptive fuzzing approach. Jiang et al.~\cite{jiang2018contractfuzzer} generate valid test inputs based on smart contract ABI specifications and monitor the EVM to detect vulnerabilities.

\subsection{Deep Learning-based Detection Methods}
In recent years, deep learning has shown great potential in the field of vulnerability detection. 
Well-trained deep learning models can learn complex program structures and syntax rules, thereby achieving high accuracy and detection effectiveness.
In the realm of vulnerability detection, deep learning methods can address the shortcomings of traditional detection methods.
For instance, 
Ashizawa et al.~\cite{ashizawa2021eth2vec} propose the Eth2Vec, utilizing neural networks to learn susceptible features from EVM bytecode and detecting vulnerabilities by comparing the similarity between target EVM bytecode and vulnerable bytecode. 
Wang et al.~\cite{wang2020contractward} design the automated vulnerability detection tool ContractWard, which is capable of detecting five types of vulnerabilities, including timestamp vulnerabilities, reentrancy vulnerabilities, arithmetic overflow, call stack vulnerabilities, and transaction order defects. 
Furthermore, Liu et al.~\cite{liu2021smart} model smart contract graphs for vulnerability detection. 
Liang et al.~\cite{liang2024ponziguard} proposed PonziGuard, which detects Ponzi contracts by combining control flow, data flow, and execution behavior information in contract behavior running graphs.

Although smart contract vulnerability detection methods based on deep learning demonstrate significant advantages,
they still face some challenges and limitations. 
One of them is the high dependence on high-quality labeled data, which is often difficult to obtain in practice. 
Besides, existing methods typically start vulnerability detection from one view of source code auditing or bytecode analysis, failing to fully utilize comprehensive information from both aspects.

\section{Dataset}
\label{Dataset}
During data collection, we analyze multiple smart contract vulnerability datasets and find that the type labels for vulnerabilities are scarce. Additionally, these datasets often do not fully cover different contract versions. For example, reentrancy vulnerabilities are mostly found in versions above 0.8, but current datasets largely do not include contracts from version 0.7 and above. To address this, we have collected and integrated several smart contract datasets from the open-source community to alleviate the issue of insufficient version coverage in existing datasets, which helps in training models to better detect and adapt to new and more complex vulnerability scenarios.

\subsection{Data Collection}
\begin{figure}[t] 
    \centering 
    \includegraphics[width=0.8\textwidth]{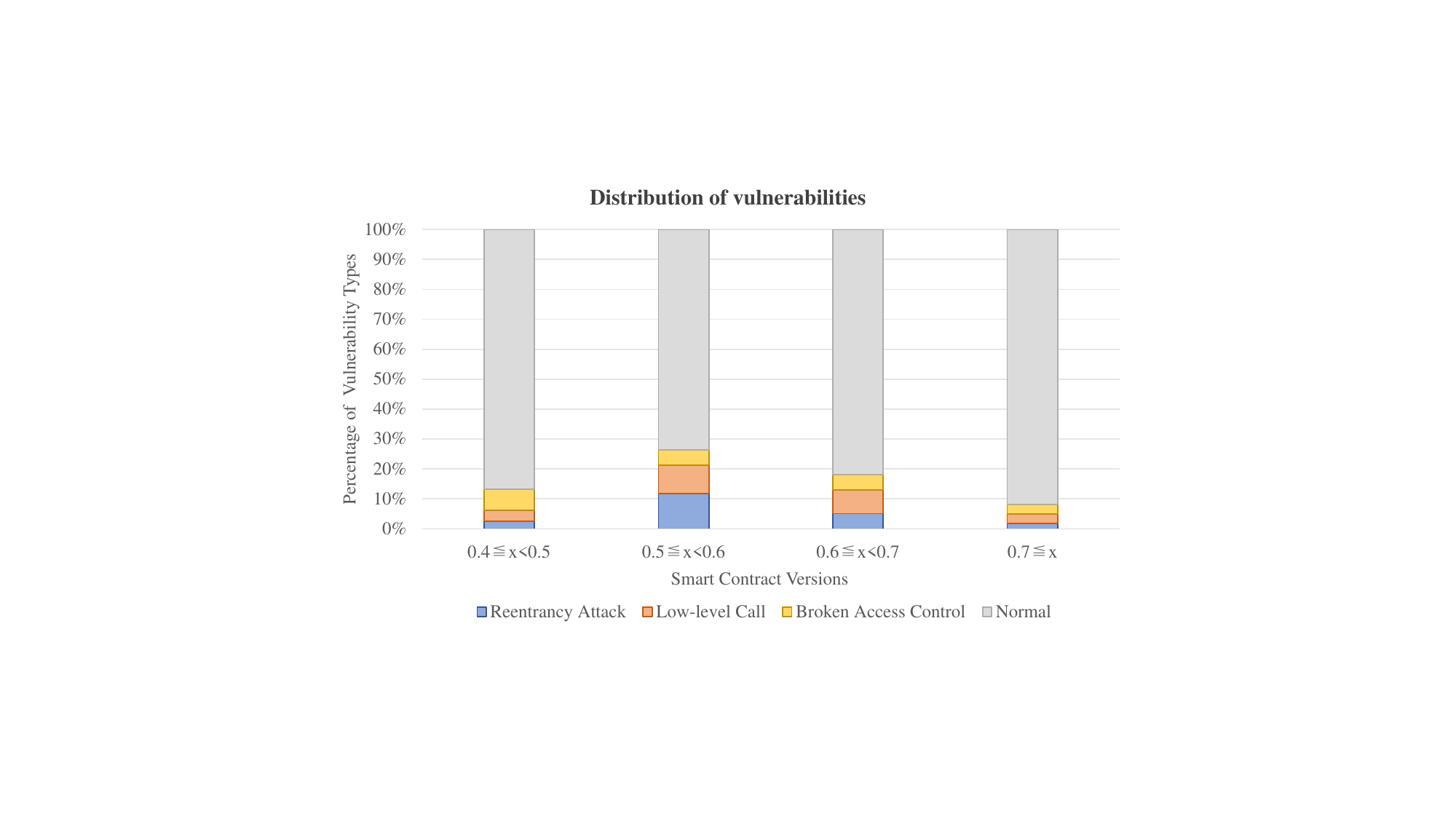} 
    \caption{Distribution of vulnerabilities in different versions of contracts.} 
\label{fig: distribution}

\end{figure}

We collect training datasets of smart contracts from Github. 
Initially, by merging multiple open-source datasets and removing duplicates, we obtain 53,000 smart contracts. 
Subsequently, we clean the data, including eliminating whitespace, comments, and code not conforming to the structure of Solidity's syntax, to obtain 35,000 smart contract datasets suitable for research.
To further obtain vulnerability labels for the dataset, we employ a voting tool to review each piece of code, selecting 10,000 smart contracts for the training set. 
Positive samples in the training set are derived from the 35,000 contracts confirmed to contain vulnerabilities through voting, while negative samples are randomly selected from smart contracts confirmed to be normal. 
The distribution of contract versions and corresponding vulnerabilities in the training set is illustrated in Fig.~\ref{fig: distribution}. 
For the testing set, positive samples are obtained from the \textit{smartbugs-curated}\footnote[2]{\url{https://github.com/smartbugs/smartbugs-curated}} dataset, while the number of negative samples matches that of positive samples, selected from normal smart contracts.

\subsection{Label Generation}
To ensure the accuracy and fairness of vulnerability labeling, we employ five voting tools to annotate smart contract vulnerabilities. 
These tools cover a variety of detection techniques, thus avoiding bias in the voting process towards any particular detection type, including Slither~\cite{feist2019slither}, Mythril~\cite{sharma2022survey}, Oyente~\cite{luu2016making}, Osiris~\cite{torres2018osiris}, Securify~\cite{tsankov2018securify}.
The voting tools combine two static analysis tools and three symbolic execution tools.
The utilization of diverse tools facilitates the acquisition of more comprehensive and precise labels, thereby improving the model's performance and generalization capabilities.

\section{Methodology}
\label{Methodology}
\begin{figure}[t]
    \centering 
    \includegraphics[width=1\textwidth]{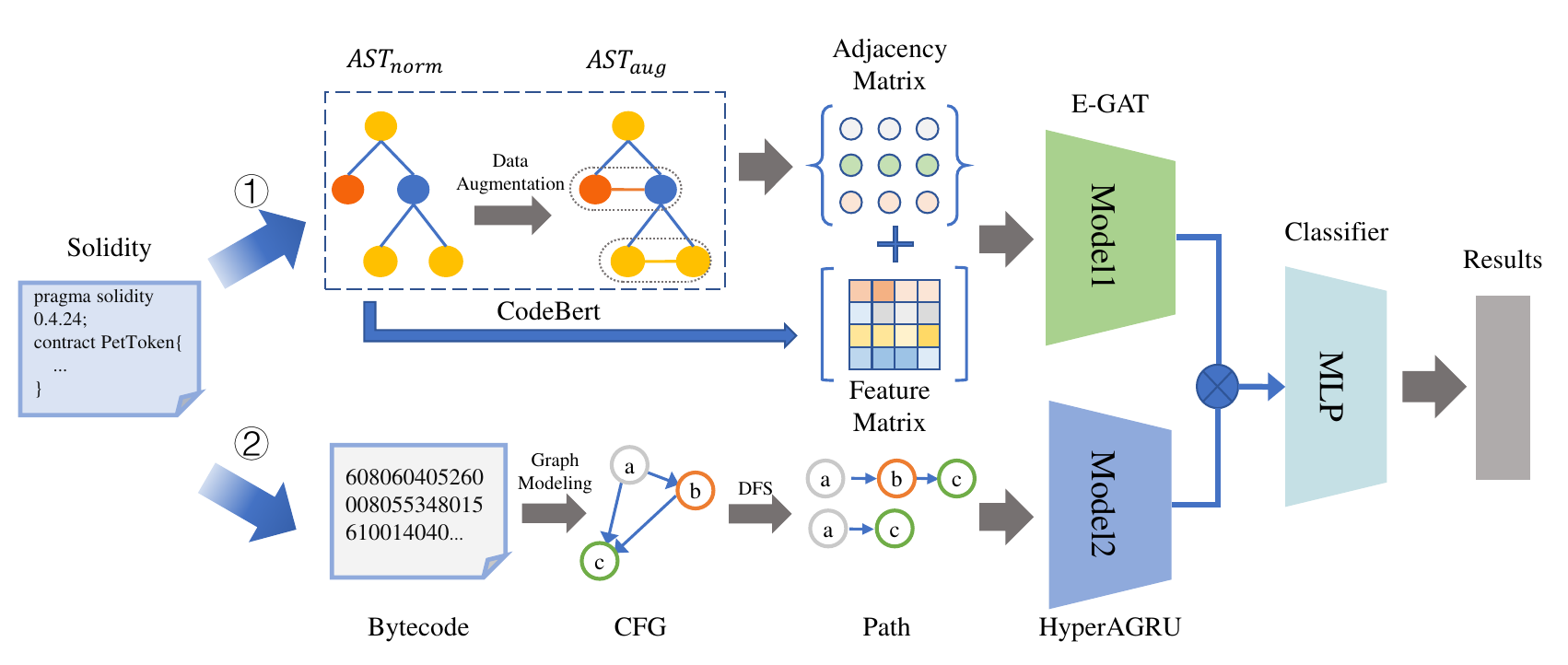} 
    \caption{Dual-view aware smart contract vulnerability detection framework.} 
\label{fig: framework}
\end{figure}

In this section, we propose a dual-view aware smart contract vulnerability detection framework (DVDet), which captures and fuses potential risk features from both the source code and bytecode of smart contracts to achieve effective detection.
The overall framework is illustrated in Fig.~\ref{fig: framework}.

\subsection{Source-code Aware Channel}
This view aims to perceive potential vulnerabilities by analyzing features of smart contract source code. 
Specifically, we first convert smart contract source code into abstract syntax trees. 
Then, we design data augmentation strategies to further transform them into weighted smart contract code graphs. 
Finally, we utilize edge-aware graph attention networks to learn node features.

\subsubsection{Augmented Smart Contract Graph Generation}
Abstract Syntax Tree (AST) is a tree-like data structure used to represent the code structure in programming languages. By analyzing the syntax and semantic structure of the code, AST converts the code into a tree representation, where each node represents a syntactic unit of the code, such as expressions, statements, or function definitions, and these nodes are interconnected via parent-child relationships. 

\begin{figure}[t] 
    \centering 
    \includegraphics[width=1\textwidth]{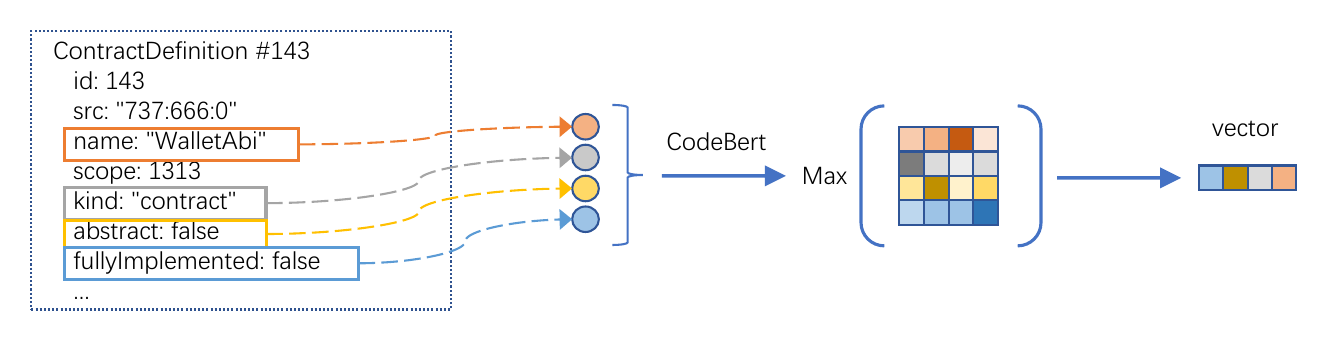} 
    \caption{Illustration of encoding AST nodes using CodeBert.} 
\label{fig: encode}
\end{figure}

In this paper, we utilize the third-party tool \textit{solc-typed-ast}\footnote[3]{\url{https://github.com/Consensys/solc-typed-ast}} to obtain normalized AST, and obtain vectorized representations of AST nodes via CodeBert~\cite{feng2020codebert}, as illustrated in Fig.~\ref{fig: encode}.
When processing AST, we focus on extracting key information from the nodes to improve the efficiency and accuracy of the analysis.
For instance, the \textit{ContractDefinition} node may contain information that is irrelevant for detection purposes. 
Hence, we retain only crucial fields such as \textit{name}, \textit{kind}, \textit{abstract}, and \textit{fullyImplemented}. 
Here, \textit{name} represents the contract name, \textit{kind} is used to distinguish whether the contract is of contract, library, or interface, \textit{abstract} indicates whether it is an abstract contract 
, and \textit{fullyImplemented} being True indicates that the contract has implemented at least one function. 
This approach not only optimizes the data structure but also ensures that the vulnerability detection can focus on the most critical information.

\begin{figure}[t] 
    \centering 
    \includegraphics[width=0.8\textwidth]{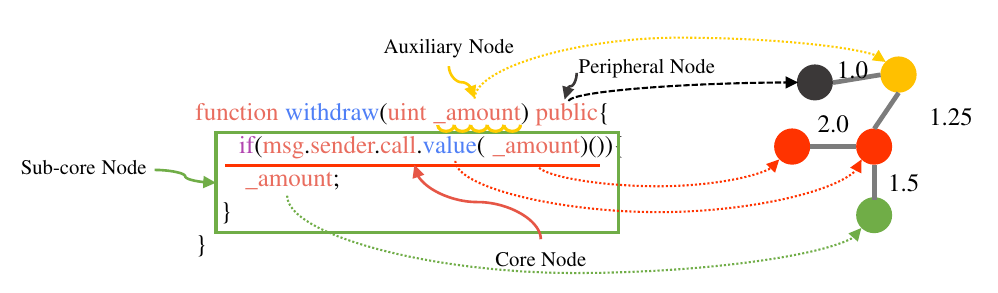} 
    \caption{An example of constructing weighted contract graph of reentry vulnerability.} 
\label{fig: example}
\end{figure}

After the source code is transformed to AST, there still exist numerous irrelevant nodes about vulnerability, which may be considered as noise and not favorable for detection. To address this issue, we propose a data augmentation method for AST, which quantifies the importance of different types of nodes to enhance the representation.
Fig.~\ref{fig: example} takes reentrancy vulnerability as an example to illustrate the augmentation process. 
On one hand, this type of vulnerability can only be triggered during execution by invoking \textit{call.value()}. On the other hand, the emergence of the vulnerability is not only related to the function caller but also to variable constraints in the context, such as certain variables triggering an \textit{if-else} branch, leading to the execution of vulnerable code. Therefore, in vulnerability detection, the \textit{if-else} blocks are more critical relative to the \textit{call.value()} node. By enhancing the edge features between \textit{if-else} and \textit{call.value()}, the influence of important neighboring nodes can be increased, allowing the neural network to focus more on crucial information during the aggregation process.

During the augmentation process, we first decompose the vulnerable contract into four types of nodes, including core nodes, secondary core nodes, auxiliary nodes, and peripheral nodes. 
Taking reentrancy vulnerability as an example, its core node is \textit{call.value()}, which directly originates the vulnerability. 
Sub-core nodes include variables \textit{require}, \textit{if}, \textit{msg.sender}, and \textit{balance}, which are relevant to the syntax structure of the core node.
Auxiliary nodes represent statements or code blocks like \textit{if-else}, \textit{while}, \textit{do-while}, \textit{for}, as well as nodes within the current function that are identical to the core and sub-core nodes. 
Peripheral nodes have minimal direct logical relationship with the core nodes and hence exert low influence on the vulnerability. 
In summary, we assign different levels of importance to these four types of nodes, denoted as $S = \{2,\ 1.5,\ 1.25,\ 1\}$.

Then, we transform the AST to a smart contract graph $G_\textit{ast} = (V, E, \boldsymbol{X})$, where $V$ represents the set of nodes in the AST, $E$ represents the set of edges formed by data flow or control flow, and $\boldsymbol{X}$ represents the feature matrix of nodes.
According to the definitions of node types mentioned above, we assign different importance weights to the edges between different node pairs as follows:
\begin{equation} \label{eq1}
    \centering
    S_\textit{ij}=\min\left(S_i,S_j\right)
\end{equation}
where $S_i$ and $S_j$ represent the importance of nodes $v_i$ and $v_j$ respectively, and $S_\textit{ij}$ denotes the importance weight of the edge. 
As a result, we transform the AST to a weighted smart contract graph $\hat{G}_\textit{ast} = (V, E, \boldsymbol{X}, S)$. 

\subsubsection{Edge-aware Attention Network}
The Graph Attention Network (GAT)~\cite{velickovic2017graph} is a prevalent graph deep learning model utilized for aggregating high-quality domain features by employing attention mechanisms. However, GAT overlooks the significance of edge features within the graph.
Therefore, we propose an Edge-aware Attention Network (E-GAT).
Firstly, based on the obtained $\hat{G}_\textit{ast}$, we regard the edge importance $S_\textit{ij}$ as edge features, which will participate in the subsequent computation of the attention:

\begin{equation}
    \label{eq2}    
    \begin{aligned}
    &\boldsymbol{e}_\textit{ij}^{(l)}=\mathrm{LeakyReLU}\left(\boldsymbol{a}^ \top\left(\boldsymbol{W}_h \boldsymbol{X}_i^{(l)}+\boldsymbol{W}_h \boldsymbol{X}_j^{(l)}\right)\cdot S_\textit{ij}\right)  \\
        &\hat{\boldsymbol{X}}_i^{(l+1)}=\sigma\left(\sum_{j\in \mathcal{N}_i}\mathrm{Softmax}\left(\boldsymbol{e}_\textit{ij}^{(l)}  \right)\cdot \boldsymbol{W}_h \boldsymbol{X}_j^{(l)}\right)
    \end{aligned}
\end{equation}

where $\boldsymbol{a}$ is the learnable attention projection vector, $\boldsymbol{W}_h$ is the learnable weight matrix for nodes and edges respectively, $\mathcal{N}_i$ is the neighbor set of node $v_i$, and $\sigma$ is the activation function. 
By aggregating the node pairs and then multiplying with the edge features, we obtain the attention coefficients $e_{ij}$. 

With the introduction of attention weights, we can selectively aggregate neighborhood information to the target node, thereby obtaining unique and refined node representations.

\subsection{Bytecode Aware Channel}
This view aims to analyze the bytecode features of smart contracts.
Specifically, we first decompile the bytecode of smart contracts into opcodes.
Then, according to jump instructions, we further transform them into a control flow graph. 
Subsequently, we utilize path-searching algorithms to obtain the control flow. 
Finally, we design sequence models HyperAGRU to learn the representations.
\begin{figure}[t] 
    \centering 
    \includegraphics[width=0.9\textwidth]{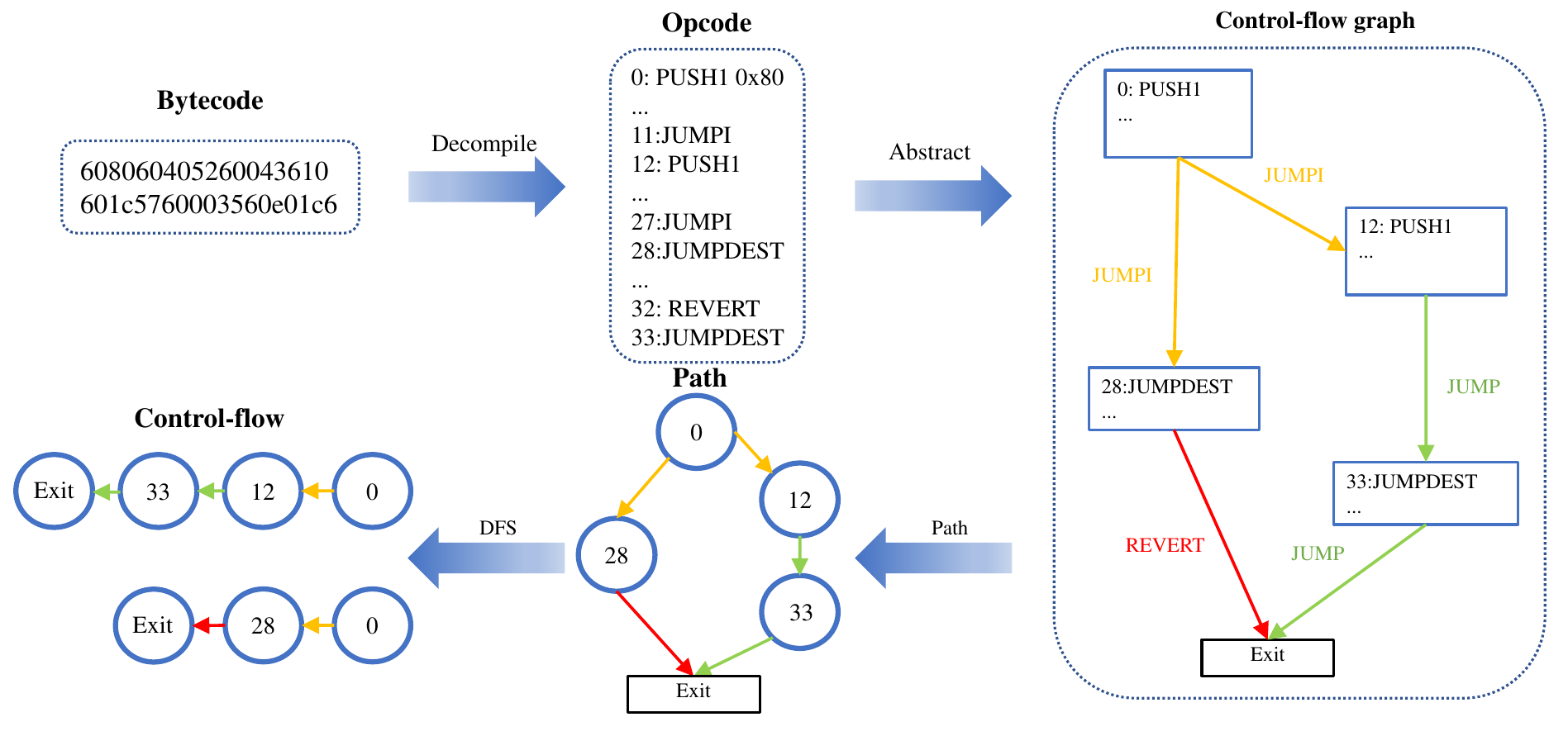} 
    \caption{Illustration of control flow generation.} 
\label{fig: flow}
\end{figure}

\subsubsection{Control Flow Generation}
Opcodes are fundamental instructions in Ethereum, which are obtained by decompiling bytecode and enable smart contracts to access memory and interact with others.
These opcodes involve accessing and modifying data~\cite{krupp2018teether} in the stack, memory, and storage devices.

We utilize these opcodes to construct the Control Flow Graph (CFG) of the contract, which is crucial for understanding the program structure and dynamics. 

In the CFG, each node represents a sequence of consecutive opcodes, known as a basic block, while edges represent the jumps between basic blocks. 

Additionally, the CFG can eliminate inactive code, reducing interference during the detection process, and providing a more detailed depiction of the data flow sequence during program execution. 
Hence, we derive all potential and crucial paths that may be traversed during program execution from the CFG.

Specifically, we acquire multiple control flows through the depth-first search algorithm,
which is illustrated in Fig.~\ref{fig: flow}.
\begin{figure}[t] 
    \centering 
    \includegraphics[width=0.5\textwidth]{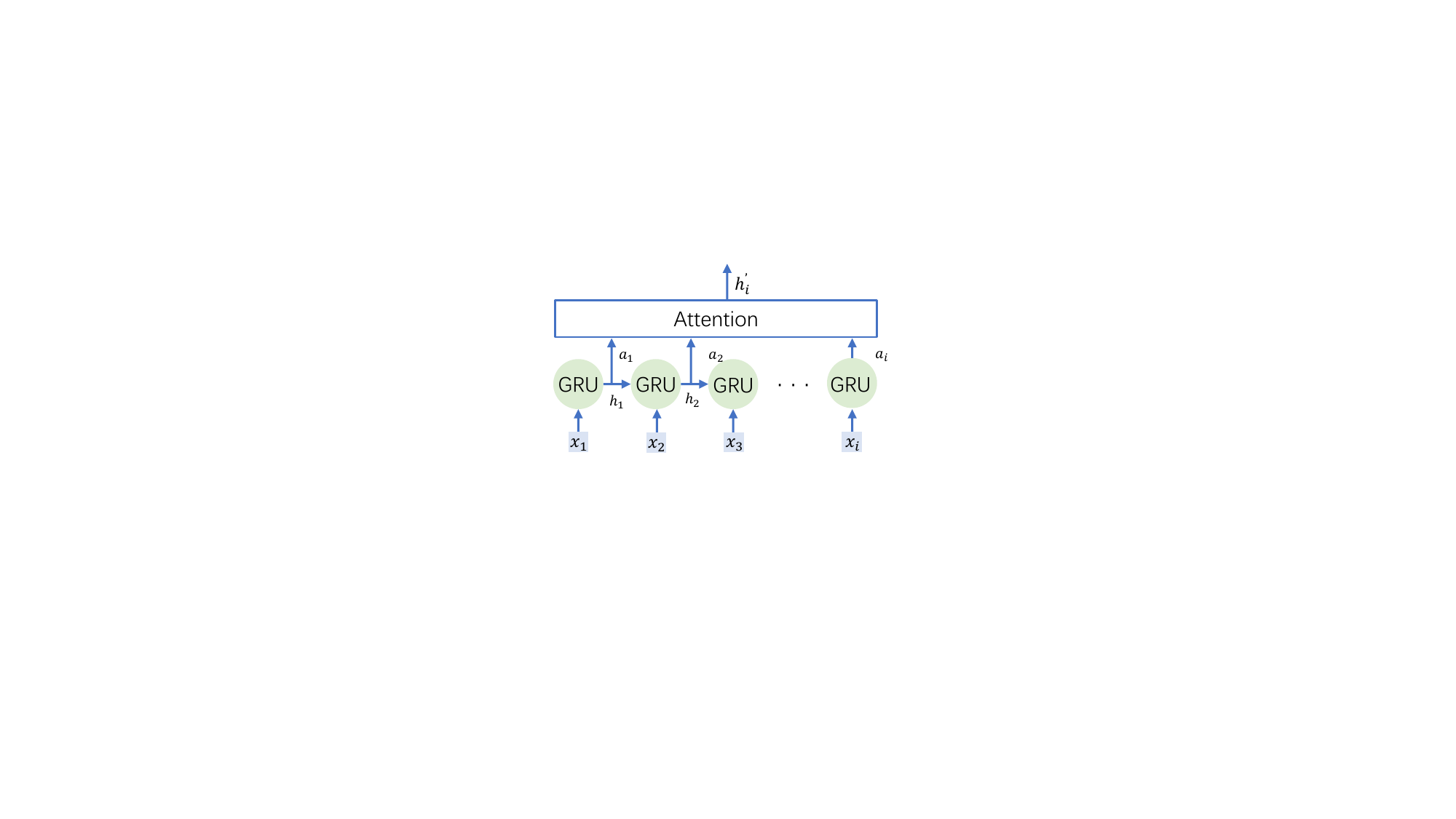} 
    \caption{The framework of the HyperAGRU.} 
\label{fig: hypergru}
\end{figure}

\subsubsection{HyperAGRU}
After obtaining the control flow data, we use sequence models to learn the feature representation of the control flow. Considering that traditional GRUs still face the problem of losing long-term dependencies when processing long sequences, we propose the HyperAGRU model. By incorporating an attention mechanism into the GRU, this model can not only capture local patterns within the control flow but also dynamically assign weights to each operation block in different contexts. This approach highlights those instructions and jumps that are crucial for security assessment.

As depicted in Fig.~\ref{fig: hypergru}, each element in the sequence is associated with an attention weight, indicating its contribution to the final aggregation. 

Multiplying each element of the sequence by its respective attention weight and summing the products yields the final representation.
Initially, the intermediate representation is computed using the GRU block, shown as follows:
\begin{equation}
    \label{eq4}
    \begin{split}
         \boldsymbol{z}_t&=\sigma\ (\boldsymbol{W}_z\cdot[\boldsymbol{h}_{t-1},\boldsymbol{x}_t]+\boldsymbol{b}_z)   \qquad 
    \tilde{\boldsymbol{h}}_{t}=\mathrm{tanh}\left(\boldsymbol{W}_{h}\cdot[\boldsymbol{r}_{t}\odot \boldsymbol{h}_{t-1},\boldsymbol{x}_{t}]+\boldsymbol{b}_{h}\right)
     \\
     \boldsymbol{r}_t&=\sigma\ (\boldsymbol{W}_r\cdot[\boldsymbol{h}_{t-1},\boldsymbol{x}_t]+\boldsymbol{b}_r) 
    \qquad
    \boldsymbol{h}_{t}=(1-\boldsymbol{z}_{t})\odot \boldsymbol{h}_{t-1}+\boldsymbol{z}_{t}\odot\tilde{\boldsymbol{h}}_{t}
    \end{split}    
\end{equation}
where $\boldsymbol{z}_t$ is the update gate to control the weight of the previous and current elements, $\boldsymbol{r}_t$ is the reset gate to control the influence of the previous element, $\widetilde{\boldsymbol{h}}_t$ is the candidate hidden state, and $\odot$ denotes element-wise product.
Finally, we obtain the intermediate representation $h_t$ for each element.

Utilizing an attention mechanism for $h_t$, we efficiently capture the local importance of the sequence and integrate this information into the final representation:
\begin{equation}
    \label{eq5}
    \begin{aligned}
         \boldsymbol{a}_t&= \mathrm{Softmax}\ (\boldsymbol{u}^\top\cdot \boldsymbol{h}_t)  \\
         \hat{\boldsymbol{h}}_t&= \sum_{t=1}^L \boldsymbol{a}_t\cdot \boldsymbol{h}_t
    \end{aligned}
\end{equation}
where $\boldsymbol{u}$ is the learnable attention parameter, $L$ is the length of the sequence, and $\hat{\boldsymbol{h}}_t$ is the final contract representation for the downstream dual-view fusing.

\subsection{Dual-view Vulnerability Detection}

After obtaining the outputs $\hat{X}$ and $\hat{h}$ from the two views, we aggregate them and employ a MLP as a classifier to generate the predicted values:
\begin{equation} \label{eq10}
    \centering
    \boldsymbol{p}_i=\mathrm{Softmax} \left(\boldsymbol{W}(\hat{\boldsymbol{X}}_i \parallel \hat{\boldsymbol{h}}_i)+\boldsymbol{b}\right)
\end{equation}
where $\parallel$ denotes the concatenation operation. 
Then, the loss function $\mathcal{L}$ adopts the cross-entropy loss function, shown as follows:
\begin{equation}
    \mathcal{L}= -\sum_i  \boldsymbol{y}_i \cdot \log(\boldsymbol{p}_i)
\end{equation}
Notably, the function can be used for both binary and multi-class classification.

\section{Experiments}

\subsection{Parameter Settings}
In this paper, we select three categories of methods for comparison, including sequential neural networks (LSTM~\cite{gers2001long} and GRU~\cite{cho2014learning}), graph neural network (GCN~\cite{kipf2016semi}, GIN~\cite{xu2018powerful}, GraphSAGE~\cite{hamilton2017inductive} and GAT~\cite{velickovic2017graph}), and traditional static analysis method (Conkas~\cite{veloso2021conkas}).
For LSTM and GRU, we set up a two-layer structure with a fixed input dimension of 350 dimensions. 
As for GNN, we adopt a three-layer structure with an input dimension of 768 dimensions. 
To ensure the reliability of the results, all experiments are subjected to three-fold cross-validation on the dataset.
Regarding parameter optimization, we employ the Adam optimizer with an initial learning rate of 0.01. 
Additionally, to adjust the learning rate to improve training effectiveness, we utilize a cosine annealing strategy. 
A uniform dropout rate of 0.5 is set during the experiments to alleviate the risk of overfitting. 
We employ two commonly used evaluation metrics: Accuracy and Recall.

\subsection{Evaluation on Vulnerability Detection}
We evaluate the effectiveness of our method through experiments on detecting the existence and types of vulnerabilities in smart contracts. 
The former determines the presence of vulnerabilities, while the latter identifies the specific types of vulnerabilities.

Table~\ref{tab: overall} reports the results of all methods, from which we can derive the follow conclusion:
\begin{itemize}
	\item[$\bullet$] Sequence models underperform in vulnerability detection because converting source code into opcode sequences can result in the loss of information related to the source code, such as variable names, which can adversely affect detection performance. Additionally, these sequence models struggle with the forgetting issue when dealing with long sequences. In contrast, our method not only incorporates attention within the bytecode-aware view to enhance the model's capacity for handling sequence data but also includes a source code-aware view to model critical information within the code, thereby achieving better detection performance.
    \item[$\bullet$] GNN-based methods perform better than sequence models because by converting source code into graphs, they can effectively capture the complex logical relationships within the source code. However, GNN-based approaches are highly dependent on effective graph construction strategies and the expressive power of initial features. In contrast, our method, by augmenting the code graph, further strengthens the semantic associations between code elements, thus achieving better detection results.
    \item[$\bullet$] The Conkas tool performs poorly in both tasks. As a static analysis tool, it tends to miss vulnerabilities that are only triggered under specific conditions. Moreover, Conkas struggles to handle highly abstract or novel programming constructs, which may introduce vulnerabilities that are difficult to detect.
\end{itemize}

\begin{table}[t]
\centering
\renewcommand{\arraystretch}{1.1} 
\caption{Accuracy of smart contract vulnerability detection. 
The bolding indicates the best result.}
\resizebox{0.9\textwidth}{!}{
\begin{tabular}{cc|cc|cccc|c|c} 
\hline\hline
\multicolumn{2}{c|}{Model}            & \ LSTM  & \ GRU \   & \ GCN \   & \ GIN \   & SAGE  & \ GAT \   & \ Conkas \ & \ DVDet  \\ 
\hline
\multirow{2}{*}{Acc (\%)} & Type      & 65.35 & 72.54 & 80.07 & 79.99 & 78.47 & 81.62 & 61.97  & \textbf{84.50}  \\
                          & Existence & 71.46 & 84.25 & 86.04 & 87.83 & 87.95 & 89.00 & 78.48  & \textbf{90.74}  \\
\hline\hline
\end{tabular}}
\label{tab: overall}
\end{table}

\begin{table}[t]
\centering
\renewcommand{\arraystretch}{1.2}
\caption{Performance of ablation experiments for different views. The bolding indicates the best results.}
\resizebox{0.9\textwidth}{!}{
\begin{tabular}{cc|cccccc|c} 
\hline\hline
\multicolumn{2}{c|}{\multirow{2}{*}{Methods}} & \multicolumn{6}{c|}{Type}                                                                           & \multirow{2}{*}{Existence}  \\
\multicolumn{2}{c|}{}                         & \multicolumn{2}{c}{ReEn}        & \multicolumn{2}{c}{LoWc}        & \multicolumn{2}{c|}{AcCl}       &                             \\
SC    & BT                                    & ~ ~Acc~ ~      & ~Recall~       & ~ ~Acc~ ~      & ~Recall~       & ~ ~Acc~ ~      & ~Recall~       & Acc                         \\ 
\hline
GAT   & ~-                                    & 84.05          & 73.91          & 81.21          & 80.55          & 78.59          & 66.66          & 81.31                       \\
E-GAT & -                                     & 85.51          & 80.63          & 82.52          & 87.62          & 78.61          & 74.75          & 82.25                       \\
-     & GRU                                   & 78.21          & 75.12          & 78.39          & 79.33          & 79.53          & 76.33          & 78.71                       \\
-     & HyperAGRU                             & 79.96          & 74.22          & 78.41          & 79.62          & 80.37          & 77.85          & 79.58                       \\
GAT   & HyperAGRU                             & 84.09          & 78.86          & 82.47          & 80.55          & 80.22          & 84.09          & 82.26                       \\
E-GAT & GRU                                   & 87.44          & 79.59          & 82.45          & 86.42          & \textbf{80.78} & 75.00          & 83.55                       \\
\multicolumn{2}{c|}{DVDet}                    & \textbf{88.73} & \textbf{82.60} & \textbf{84.50} & \textbf{88.66} & 80.28          & \textbf{87.44} & \textbf{84.50}              \\
\hline\hline
\end{tabular}
}
\label{table2}
\end{table}

\subsection{Ablation Experiment}
To evaluate the effectiveness of our dual-view framework and our improved models, we perform a series of ablation studies, as shown in Table~\ref{table2}. Specifically, SC indicates source code view, and BT indicates the bytecode view. 

We can derive the following conclusions: 
\begin{itemize}
	\item[$\bullet$] From the source code view, our E-GAT outperforms GAT, indicating that our data augmentation strategy designed for contract code graphs effectively strengthens the associations between code elements, aiding in the enhancement of vulnerability detection;
    \item[$\bullet$] From the bytecode view, our HyperAGRU generally outperforms GRU, demonstrating that incorporating attention mechanisms into sequence encoding can effectively capture the rich semantic information of the code flow, thereby enhancing vulnerability detection;
    \item[$\bullet$] By comparing DVDet with E-GAT+SC (or HyperAGRU+BT), it is evident that there is a complementary effect between the two perspectives within the DVDet framework.
\end{itemize}

\begin{table}[t]
\centering
\setlength{\tabcolsep}{4pt}
\renewcommand{\arraystretch}{1.3}
\caption{Efficiency of Smart Contract Vulnerability Detection.}
\resizebox{0.9\textwidth}{!}{
\begin{tabular}{ccccccc} 
\hline\hline
Methods                                                    & Slither & Securify & Oyente & Osiris  & Mythril & DVDet           \\ 
\hline
\textcolor[rgb]{0.173,0.173,0.212}{Unit Detection Time(s)} & 2.09    & 59.17    & 4.51   & 21.16   & 31.64   & \textbf{0.14}   \\
Total Detection Time(s)                                    & 409.64  & 10058.92 & 816.31 & 4147.36 & 6201.44 & \textbf{27.17}  \\
\hline\hline
\end{tabular}}
\label{tab: time}
\end{table}
\subsection{Efficiency Experiment}
To evaluate the efficiency of our method in vulnerability detection, we conduct efficiency analysis experiments. The results, as shown in Table~\ref{tab: time}, indicate that tools based on symbolic execution, such as Securify, Osiris, and Mythril, have a large time consumption on vulnerability detection. Slither and Oyente demonstrate faster detection speeds. Our DVDet framework exhibits exceptionally high efficiency in vulnerability detection, achieving an order of magnitude advantage over some existing detection tools.

\section{Conclusion} \label{conclusion}
Smart contract vulnerabilities have caused serious damage to the ecosystem of the Ethereum platform. 
Existing methods typically adopt a single view when designing algorithms for vulnerability detection, focusing solely on either the source code perspective or the bytecode perspective. 
In this paper, to address the issue of one-sided views in existing methods, we propose a smart contract vulnerability detection method that incorporates dual-view fusion. 

Extensive experiments demonstrate that our framework achieves outstanding detection performance, indicating that the dual view can acquire more comprehensive information for vulnerability detection. 
In addition, compared with other traditional detection methods, the deep learning-based method also demonstrates significantly higher detection efficiency, offering a new perspective for subsequent smart contract vulnerability detection.

\subsubsection*{Acknowledgments.} 
This work was supported in part by the Key R\&D Program of Zhejiang under Grants 2022C01018 and 2024C01025, by the National Natural Science Foundation of China under Grants 62103374 and U21B2001.

\bibliographystyle{splncs04_}
\bibliography{blocksys-DVDet1}

\end{document}